\documentclass[twocolumn,twoside,5p,preprint]{elsarticle}
\biboptions{sort&compress}

\usepackage[utf8]{inputenc}
\usepackage[english]{babel}
\usepackage{amssymb,amsthm,amsmath,amstext,amsbsy,amsopn}
\usepackage{bbm}
\usepackage{graphicx}
\usepackage{isotope}
\usepackage{float}
\usepackage[dvipsnames]{xcolor}
\usepackage[T1]{fontenc}

\usepackage{hyperref}
\hypersetup{
    colorlinks = true,
    citecolor  = blue,
    linkcolor  = blue,
    urlcolor  = black
}

\newcommand{\MeV}{\text{MeV}}

\newcommand{\ra}{\rangle}

\newcommand{\Efourptwop}{\ensuremath{E^\star_{4^+}/E^\star_{2^+}}}

\newcommand{\ai}{{\emph{ab initio}}}
\newcommand{\eg}{{\emph{e.g.}}}
\newcommand{\ie}{\textit{i.e.}}
\newcommand{\etal}{\textit{et al.}}
\newcommand{\hilbert}[1]{\mathcal{H}_{#1}}

\newcommand{\Xmax}[1]{#1_\text{max}}

\newcommand{\trace}{\text{Tr}}

\newcommand{\NNN}{\text{3N}}
\newcommand{\elem}[2]{$^{#1}$#2}

\begin{document}

\title{Spectroscopy of $N=50$ isotones with the \\
valence-space density matrix renormalization group}

\address[tud]{Technische Universit\"at Darmstadt, Department of Physics, 64289 Darmstadt, Germany}
\address[emmi]{ExtreMe Matter Institute EMMI, GSI Helmholtzzentrum f\"ur Schwerionenforschung GmbH, 64291 Darmstadt, Germany}
\address[mpik]{Max-Planck-Institut f\"ur Kernphysik, Saupfercheckweg 1, 69117 Heidelberg, Germany}
\address[wigner]{Wigner Research Centre for Physics, P.O.~Box 49, H-1525 Budapest, Hungary}
\address[dtpbudapest]{Department of Theoretical Physics, Institute of Physics, Budapest University of Technology and Economics, M\H uegyetem rkp.~3, H-1111~Budapest, Hungary}
\address[tu]{Center for Computational Sciences, University of Tsukuba, 1-1-1 Tennodai, Tsukuba, Ibaraki 305-8577, Japan}
\address[mtabme]{MTA-BME Quantum Dynamics and Correlations Research Group, Budapest University of Technology and Economics, M\H uegyetem rkp.~3, H-1111~Budapest, Hungary}
\address[ias]{Institute for Advanced Study, Technical University of Munich, Lichtenbergstrasse 2a, 85748 Garching, Germany}

\author[tud,emmi,mpik]{A.\ Tichai}
\ead{alexander.tichai@physik.tu-darmstadt.de}

\author[wigner,dtpbudapest]{K.\ Kap\'as}
\ead{kapas.kornel@wigner.hun-ren.hu}

\author[tud,emmi,mpik,tu]{T.\ Miyagi}
\ead{miyagi@physik.tu-darmstadt.de}

\author[mtabme,dtpbudapest,wigner]{M.\ A.\ Werner}
\ead{werner.miklos@ttk.bme.hu}

\author[wigner,ias]{\"O.\ Legeza}
\ead{legeza.ors@wigner.hu}

\author[tud,emmi,mpik]{A.\ Schwenk}
\ead{schwenk@physik.tu-darmstadt.de}

\author[dtpbudapest,mtabme]{G.\ Zarand}
\ead{zarand.gergely.attila@ttk.bme.hu}

\begin{abstract}
The recently proposed combination of the valence-space in-medium similarity renormalization group (VS-IMSRG) with the density matrix renormalization group (DMRG) offers a scalable and flexible many-body approach for strongly correlated open-shell nuclei. We use the VS-DMRG to investigate the low-lying spectroscopy of $N=50$ isotones, which are characteristic for their transition between single-particle and collective excitations. We also study electromagnetic transitions and show the advantage of the VS-DMRG to capture the
underlying physics more efficiently, with significantly improved convergence compared to state-of-the-art shell-model truncations.
Combined with an analysis of quantum information measures, this further establishes the VS-DMRG as a valuable method for \ai{} calculations of nuclei.
\end{abstract}

\maketitle

\section{Introduction}

The first-principles solution of the nuclear many-body problem has witnessed continuous advances to heavier systems and exotic nuclei at the neutron-rich extremes.
This success is driven by developments in quantum many-body theory (see, \eg, Refs.~\cite{Herg20review,Hebe203NF}) and from nuclear forces derived within chiral effective field theory (EFT)~\cite{Epel09RMP,Mach11PR,Hamm20RMP}.
With this it became possible to accurately describe more than one hundred fully interacting nucleons in a controlled way~\cite{Morr17Tin,Arthuis2020a,Hu2021lead,Hebeler2023jac,Tichai2024bcc,Arthuis2024}.
In medium-mass nuclei, the many-body solution is commonly obtained through expansion methods that scale polynomially with system size, \eg{}, using coupled cluster (CC) theory~\cite{Hage14RPP}, the in-medium similarity renormalization group (IMSRG) approach~\cite{Tsuk11IMSRG,Herg16PR,Heinz2020,Stroberg2019}, Green's function theory~\cite{Dick04PPNP,Soma20SCGF}, or many-body perturbation theory (MBPT)~\cite{Tichai2020review}.

In their basic closed-shell formulation, the many-body expansion is performed around a spherical Hartree-Fock (HF) reference state that is obtained through the solution of the symmetry-restricted HF equations. When targeting nuclei far away from shell closures, including static correlations becomes challenging in this basic expansion.
To this end, valence-space methods provide a powerful set of tools to obtain an accurate description of open-shell nuclei.
Here the initial Hamiltonian is mapped onto a suitably chosen valence space with only a small number of active nucleons that can be explored through standard diagonalization techniques.
In the context of \ai{} calculations the construction of effective valence-space operators has been initially performed using MBPT (see, \eg, Ref.~\cite{Holt14Ca} based on modern nuclear forces). More recently, the valence space approach was developed for CC and IMSRG frameworks~\cite{Tsukiyama2012,Bogn14SM,Jans14SM,Stro17ENO,Sun2018,Stroberg2019,Stroberg2021}.
While the final valence-space Hamiltonian diagonalization is technically identical to a shell-model calculation~\cite{Caur05RMP}, this provides a first-principles construction of the valence-space Hamiltonian based on chiral EFT.
Still valence-space approaches strongly benefit from the methodological advances for solving large-scale eigenvalue problems~\cite{kshell,Johnson2018bigstick}.
Alternatively, open-shell nuclei are commonly tackled through multi-reference techniques (see, \eg, Refs.~\cite{Herg13MR,Gebr17IMNCSM,Yao18IMGCM,Tich17NCSM-MCPT,Frosini2021mrI,Frosini2021mrII,Frosini2021mrIII}) or symmetry-broken expansions~\cite{Soma13GGF2N,Tichai18BMBPT,Novario2020a,Demol20BMBPT,Hagen2022PCC,Tichai2024bcc}.

However, even the most advanced shell-model solvers cannot cope with the dimensions of the Hilbert spaces that emerge in large-scale applications.
We recently proposed the merging of the density-matrix renormalization group (DMRG) with the VS-IMSRG as a scalable alternative for the description of low-lying states in strongly correlated nuclei~\cite{Legeza2015,Tichai2023dmrg}.
It was demonstrated that energies can be converged much more rapidly within the VS-DMRG approach, thus, reducing the associated many-body uncertainty.
With the resulting VS-DMRG we were further able to successfully reproduce the experimentally observed $N=50$ shell closure in \elem{78}{Ni}~\cite{Taniuchi2019} and link the nuclear shell structure to a drop in total entanglement in the many-body wave-function, revealing an interesting connection between nuclear phenomenology and quantum information (QI) measures.

To put the VS-DMRG approach on par with conventional diagonalization techniques, however, one must be able to describe the same set of nuclear observables.
Hence in this work we extend the VS-DMRG approach to electromagnetic transition operators needed for the extraction of $B(E2)$ values. With this at hand, we explore the $N=50$ isotones that are governed by a rapid transition between single-particle and collective excitations~\cite{Nowacki2016,Li2023}. Moreover, we show the advantage of the VS-DMRG to capture the
underlying physics more efficiently and exemplify this improvement using QI measures.

\section{Methodology}

For the VS-DMRG calculations, we employ the VS-IMSRG to calculate from a (free-space) nuclear Hamiltonian $H$ an effective valence-space Hamiltonian. This evolution is mediated by a unitary transformation $U(s)$ controlled by a flow parameter $s$, $H(s) = U^\dagger(s) H(0) U(s)$.
The VS-IMSRG flow nonperturbatively decouples the valence space to lower and higher lying particle and hole excitations~\cite{Herg16PR,Stroberg2019}.
Practically, the VS-IMSRG transformation yields a set of second-quantized valence-space operators
\begin{align}
    O = o_0 + \sum_{pq} o_{pq} \, c^\dagger_p c_q 
    + \frac{1}{4} \sum_{pqrs} o_{pqrs} \, c^\dagger_p c^\dagger_q c_s c_r \, ,
    \label{eq:op}
\end{align}
where $o_0$, $o_{pq}$, and $o_{pqrs}$ denote the normal-ordered \text{zero-,} one- and (anti-symmetrized) two-body matrix elements, respectively. The single-particle state $|p \ra = | n l j m_j m_t \ra$ gathers all quantum numbers: radial quantum number $n$, orbital angular momentum $l$, total angular momentum $j$ with projection $m_j$, and isospin projection $m_t$ distinguishing protons and neutrons.
The commutator evaluation within the VS-IMSRG transformation naturally induces high-body operators that are truncated at the normal-ordered two-body level, while discarding operators at rank three and beyond~\cite{Herg16PR,Heinz2020}.
The estimated truncation effect on ground-state observables is of the order of $1-2\%$~\cite{Heinz2020}.
In practice, we employ the Magnus formulation of the IMSRG which enables solving for the transformation itself instead of the transformed Hamiltonian~\cite{Morr15Magnus}.
We highlight that the IMSRG evolution requires a consistent transformation of other operators as well, \eg{}, for electromagnetic transition operators.
While the leading multipole operators are of one-body character, the commutator evaluation induces non-scalar higher-body components that need to be included explicitly.

Once the VS-IMSRG decoupling is performed, the valence-space Hamiltonian is used as input for a variational density matrix renormalization group (DMRG) calculation~\cite{White1992,Schollwoeck2011,Szalay2015tensor}. The fermionic single-particle orbitals in Eq.~\eqref{eq:op} are mapped onto a one-dimensional tensor topology in occupation-number representation with local dimension $d=2$ (labeled by $\sigma_p \in \lbrace 0,1 \rbrace$ referring to unoccupied/occupied orbitals, respectively). The length of the chain is given by the number of single-particle states in the valence space $N = \dim \, \mathcal{H}_1$.
Details of the mapping procedure are discussed in Sec.~\ref{sec:spinchain}.
After the lattice mapping the wave function is represented in a matrix-product-state (MPS) representation.
The corresponding wave function of $N$ orbitals is an $N$-dimensional tensor, and the CI coefficient corresponding to a determinant $\boldsymbol{\sigma}=(\sigma_1,\sigma_2,\ldots, \sigma_p,\sigma_{p+1},\ldots,\sigma_N)$ is expressed as a product of matrices $A_p^{\sigma_p}$ associated to each orbital $p$ as  $| \Psi \rangle = \sum_{\boldsymbol{\sigma}}C_{\boldsymbol{\sigma}}|\boldsymbol{\sigma}\rangle$, 
where
\begin{align}
   C_{\boldsymbol{\sigma}} = 
   A_1^{\sigma_1} 
   A_2^{\sigma_2} 
   \ldots 
   A_p^{\sigma_p} 
   A_{p+1}^{\sigma_{p+1}} 
   \ldots 
   A_N^{\sigma_N}\,.
    \label{eq:mps}
\end{align}
The size of the MPS tensor would scale exponentially with system size. Therefore, in practice, truncations are employed by setting an upper limit $M$ (the so-called bond dimension) on the number of singular values kept, and hence the maximum size of the MPS factors $A_p^{\sigma_p}$. The memory cost of storing the largest MPS factors scales as $\sim d M^2$. Our DMRG implementation exploits Abelian symmetries $G_\text{sym} =U_N(1) \, \times \, U_Z(1) \, \times \, U_{J_z}(1) \, \times \, \mathbb{Z}_2$, related to neutron-/proton-number conservation, axial symmetry and parity conservation, thus, yielding block-diagonal MPS tensors with respect to the associated quantum numbers~\cite{dmrg-budapest}.

Finally, the MPS components $A_p^{\sigma_p}$ are variationally optimized in the DMRG algorithm through iterative least-square updates. 
The tensor space is split according to $\hilbert{}^{N} = \hilbert{}^{(\text{left})} \,\otimes\, \hilbert{p} \,\otimes\, \hilbert{p+1} \,\otimes\, \hilbert{}^{(\text{right})}$ where $\hilbert{}^{(\text{left})}$ ($\hilbert{}^{(\text{right})}$) denote the left (right) blocks that are formed from precontracted $A$ matrices to the left and right of the sites $p$ and $p+1$, respectively. 
The DMRG is then executed for a sequence of bond dimensions, and the converged result is obtained by either extrapolating $1/M \rightarrow 0$ or as a function of the truncation error given by the discarded singular values~\cite{Schollwoeck2005,Legeza1996accuracy}. In this work, the latter procedure has been used by fixing the truncation error and dynamically adapting the $M$ value above a minimum threshold of $M>1024$~\cite{Legeza2003controlling}.

The initial VS-IMSRG decoupling is performed in a single-particle space comprising 15 major harmonic-oscillator shells, \ie{}, $\Xmax{e} = \text{max}(2n +l) = 14$, and the \NNN{} interaction matrix elements are restricted to $e_1 + e_2 + e_3 \leqslant E_{3\text{max}} = 24$ to ensure convergence in heavier systems~\cite{Miyagi2021}.
For all our calculations, we employ the ``1.8/2.0 (EM)'' NN+3N Hamiltonian from Ref.~\cite{Hebe11fits}, which is based on chiral EFT nucleon-nucleon (NN) and three-nucleon (3N) interactions.
The 3N contributions are approximately taken into account by keeping only up to two-body operators after normal-ordering~\cite{Hage07CC3N,Roth12NCSMCC3N,Herg13IMSRG}.
For our study of $N=50$ isotones we use a $0\hbar \omega$ valence space built on top of a \elem{60}{Ca} core, \ie{}, the proton $pf$ shell and neutron $sdg$ shell. The resulting valence space consists of 50 single-particle states (20 protons and 30 neutrons).

\section{VS-DMRG vs.~CI many-body convergence}

\begin{figure}[t!]
    \centering
    \includegraphics[width=\columnwidth]{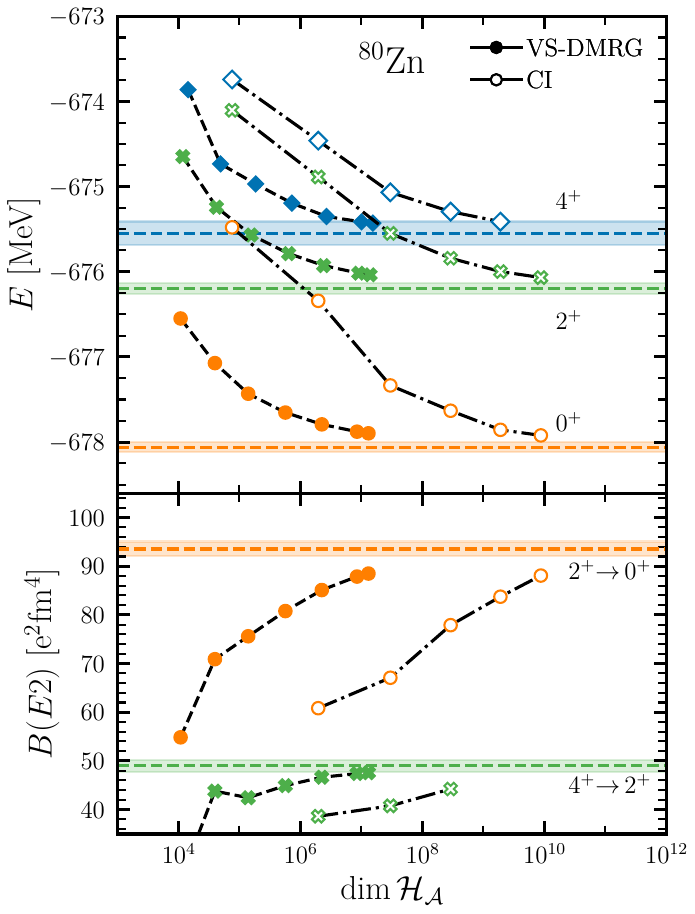}
    \caption{Convergence of VS-DMRG and CI energies (top panel) and $B(E2)$ transition strengths (bottom panel) of the lowest $0^+$, $2^+$, and $4^+$ states and their transitions as a function of Hilbert-space dimension for \elem{80}{Zn}. The largest CI dimension encountered at $\Xmax{T}=7$ is given by $\dim \mathcal{H}_A = 10^{9.94}=8.8\times 10^{9}$. The VS-DMRG extrapolated energies and transition strengths (and their uncertainties) are shown by the horizontal bands, see text for details.}
    \label{fig:conv}
\end{figure}

Next we study the improved convergence of the DMRG wave function due to its more efficient encoding of quantum correlations in the MPS ansatz compared to the linear CI expansion. For large valence spaces in shell-model calculations, where a full CI solution is not possible, one usually relies on further particle-hole truncations to make the CI diagonalization possible. Therefore, to compare the DMRG truncation in terms of bond dimensions with the particle-hole excitation level $T_\text{max}$, we use the dimension of the effective Hilbert space as a common measure to quantify performance.
In Fig.~\ref{fig:conv} we show the ground-state energies of the lowest $0^+, 2^+$, and $4^+$ states (top panel) and the allowed $B(E2)$ transition strengths (bottom panel) for the case of \elem{80}{Zn} ($Z=30$).
The largest VS-DMRG run was performed at bond dimension $M=10240$, whereas the CI calculations are carried out up to $\Xmax{T}=7$, \ie{}, including up to $7$-particle-$7$-hole excitations from the energetically lowest configuration.
For all states the VS-DMRG convergence is much faster than for CI and comparable accuracy in energy is obtained at Hilbert space sizes that are reduced by two orders of magnitude.  
We employ an extrapolation scheme based on taking the largest truncation error $\epsilon$ of the last sweep and using a linear fit to obtain the limit $\epsilon \rightarrow 0$~\cite{Legeza1996accuracy,Legeza2003controlling,Wouters2014density,Friesecke2023}.
Extrapolation uncertainties are obtained by varying the included data points in the fitting procedures.
They are depicted by horizontal bands in Fig.~\ref{fig:conv}.

Similarly, electromagnetic transitions converge more rapidly in the VS-DMRG approach as compared to the CI counterpart.
Since the $B(E2)$ values are not variational, convergence is not necessarily monotonic. While for the $2^+\rightarrow 0^+$ transition the results still reveal a monotonic increase, this is not true for the $4^+\rightarrow2^+$ transition.
To provide further support of the extrapolation, we have repeated the VS-DMRG calculations for the \elem{80}{Zn} for a set of fixed truncation errors in the range of $10^{-4}$ and $10^{-5}$. 
Including the last $3-7$ data points corresponding to the smallest truncation errors 
(with bond dimensions up to $M\simeq 13000$)
we obtain
$E_{0^+}=-678.06(6) \, \MeV$, $E_{2^+}=-676.20(7) \, \MeV$, and $E_{4^+}=-675.60(8) \, \MeV$.
Similarly, for the electromagnetic transitions we obtain $B(E2; 2^+ \rightarrow 0^+)=93.5(15) e^2 \, \text{fm}^4$ and $B(E2; 4^+\rightarrow 2^+)=49.4(8) e^2 \, \text{fm}^4$.
Hence, the extrapolation procedure allows for a robust extraction of $B(E2)$ values in VS-DMRG simulations.

\section{Orbital ordering and DMRG tensor topology}
\label{sec:spinchain}

So far the DMRG topology, \ie{}, the mapping of nuclear orbitals onto a one-dimensional lattice has been left unspecified.
In the limit of $M \rightarrow \infty$, the ordering of sites is irrelevant, and all choices will yield the full-space results.
However, every simulation at finite bond dimension is sensitive to the employed site ordering.
Moreover, for certain topologies the DMRG algorithm may get stuck in local minima.
For model Hamiltonians with nearest-neighbor interactions, \eg{}, a one-dimensional Hubbard Hamiltonian, the ordering problem is less severe since there are no long-range interactions in the system. 
While the impact of site ordering has been extensively discussed in electronic structure applications (see, \eg, Refs.~\cite{Chan2002,Legeza2003,Rissler2006,Barcza-2011}), there is only a brief documentation for nuclear physics applications~\cite{Legeza2015}.
In general, the problem of finding the ``best'' ordering is computationally complex, and requires the testing of all $N!$ orbital arrangements in a numerical simulation. For realistic problem sizes with $N \gtrsim 20$, this is clearly impractical, and one relies on heuristic arguments of choosing the DMRG topology.
Ideally, such heuristics are guided by our microscopic understanding of the many-body correlations between the interacting particles.

\begin{figure}[t!]
    \centering
    \includegraphics[width=\columnwidth]{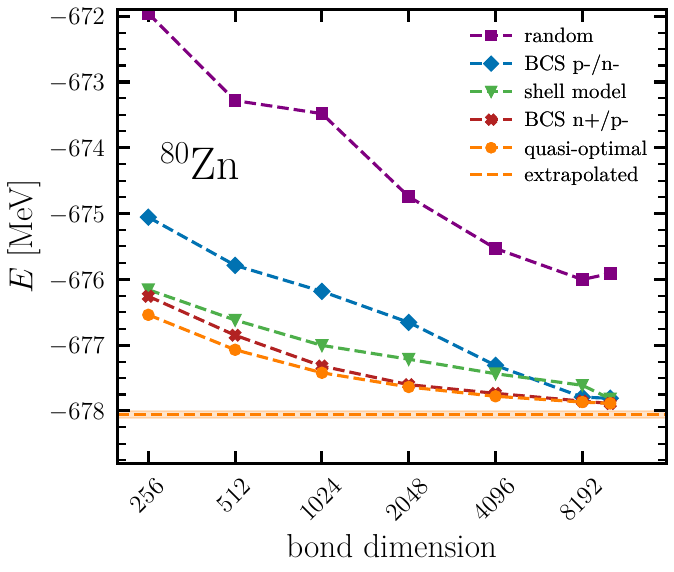}
    \caption{VS-DMRG ground-state energy of \elem{80}{Zn} as a function of bond dimension obtained for different orbital orderings in the DMRG. The different orderings correspond to: randomly sampled tensor topology with mixed protons and neutrons (purple), energetic shell-model ordering (green), two different BCS orderings (blue/red; see text for details), and quasi-optimal ordering based on single-site entropies (orange).}
    \label{fig:ordering}
\end{figure}

The efficiency of MPS encoding will be reduced in the presence of long-range correlations along the DMRG chain, \ie{}, coupling and correlations between distant sites.
In this case, the bond dimension has to be significantly increased to capture correlations. 
As a guiding principle, one must identify the dominant pairwise correlations in the system, and map the associated orbitals next to each other.
In nuclear physics such correlations manifest in the superfluid character of open-shell nuclei, \ie{}, the formation of Cooper pairs between time-reversed states due to short-range attractive interactions.
Corresponding entanglement patterns have been observed in pairwise correlation measures in open-shell nuclei~\cite{Legeza2015,Robin2021,Tichai2023dmrg,PerezObiol2023entanglement}.
In the subsequent analysis, we restrict ourselves to like-particle (pp/nn) pairs, and order states accordingly. We neglect the possibility of the formation of pn pairs that might be important in $N=Z$ nuclei.

In the DMRG the theoretical uncertainty stems from the following sources: $i)$ the accuracy threshold enforced in the diagonalization of the effective Hamiltonian, $ii)$ the truncation error related to the weight of the discarded renormalized states in the Schmidt spectrum, and $iii)$ the so-called environment error, \ie{}, the approximate representation of the left or right blocks. The final error is a function of the truncation error~\cite{Legeza1996accuracy} once the environmental error is eliminated. Since DMRG is a local optimization procedure, the ordering of orbitals has an important effect on how the environmental error is suppressed and on the required bond dimension to reach a given accuracy. 
In general, in presence of truncations the DMRG can be trapped in local minima if non-optimal orderings are employed and hence the results are more sensitive to the underlying bond dimension.

To illustrate the importance of orbital ordering,
Fig.~\ref{fig:ordering} shows the VS-DMRG convergence of the ground-state energy of \elem{80}{Zn} for a selection of DMRG topologies.
While all calculations are variationally improved with increasing bond dimension, the individual performance is vastly different.
The random ordering has a significant energetic offset compared to the other ones, converges by far the slowest in this benchmark, and may fluctuate withing the large numerical error induced by the diagonalization and environment error.
In these random arrangements we break the block arrangement of proton/neutron orbits.
Following our intuition regarding the relevance of superfluid correlations in nuclei, we next investigate two topologies with time-reversed states mapped next to each other.
The first one, labeled ``BCS p-/n-'' employs a np-block with decreasing $j$ values of the multiplets, \ie{},
\begin{align}
    \{ 
    \pi: f_{7/2} \, f_{5/2} \, p_{3/2} \, p_{1/2} \, 
    ; \,
    \nu: g_{9/2} \, g_{7/2} \, d_{5/2} \, d_{3/2} \, s_{1/2}
    \} \,, 
\end{align}
whereas the second one, labeled ``BCS n+/p-'' employs a np-block with increasing neutron orbitals and decreasing proton orbitals, \ie{},
\begin{align}
    \{ 
    \nu: s_{1/2} \, d_{3/2} \, d_{5/2} \, g_{7/2} \, g_{9/2}
    ; \,
    \pi: f_{7/2} \, f_{5/2} \, p_{3/2} \, p_{1/2}
    \} \,. 
    \label{eq:order}
\end{align}
These topologies differ in their respective order of the proton and neutron sub-blocks, and also in the order of multiplets in the proton sub-block.
While already the ``BCS p-/n-'' ordering (blue) substantially improves upon the previous random ordering, a significant gain is obtained by placing the multiplets of highest angular momentum ($\pi f_{7/2}$ and $\nu g_{9/2}$) to the center of the chain (red), giving close-to-optimal results. This highlights two aspects: $i$) BCS-type pairing correlations are relevant for minimizing long-range entanglement within the DMRG topology and $ii$) placing orbitals with higher entropies (see also Sec.~\ref{sec:QIT}) to the center of the chain is more advantageous.
As another alternative, we pick the naive shell-model ordering (green) based on the energetic ordering of orbitals while forming separate proton and neutron blocks.
This ordering yields a systematic convergence pattern, but is quite far from the ``BCS n+/p-'', especially at larger bond dimensions.

For all investigated bond dimensions, the energetically lowest results were obtained from the ``quasi-optimal'' ordering (orange) by placing orbitals with largest site entropy around the center of the DMRG chain. The resulting order of multiplets is identical to the one displayed in Eq.~\eqref{eq:order}.
However, while the quasi-optimal ordering yields the same order of $j$ multiplets as the BCS n+/p- ordering it differs in the organization of the magnetic substates; the BCS ordering employs an arrangement $m_j = \frac{j}{2},-\frac{j}{2},...,\frac{1}{2},-\frac{1}{2}$, whereas in the quasi-optimal one angular-momentum projections are increasing $m_j=-\frac{j}{2},..., \frac{j}{2}$. 
The identical ordering of multiplets explains the closeness of the results already at low bond dimension.
In essence, highly entangled orbits are those close to the Fermi surface, and are centered in the chain, while highly excited orbitals are optimally mapped towards the tensor topology's boundaries.
Since it is computationally intractable to exhaust all possible arrangements, we label this ordering as ``quasi-optimal''.
Clearly, our study shows the high relevance of ordering effects in nuclear applications and demonstrates how the quality of the results is linked to the underlying microscopic structure of a nucleus.
Still, with increasing bond dimension, all orderings but the random one yield very similar energies.
 
\section{Low-lying spectroscopy of $N=50$ isotones}

\begin{figure}[t!]
    \centering
    \includegraphics[width=\columnwidth]{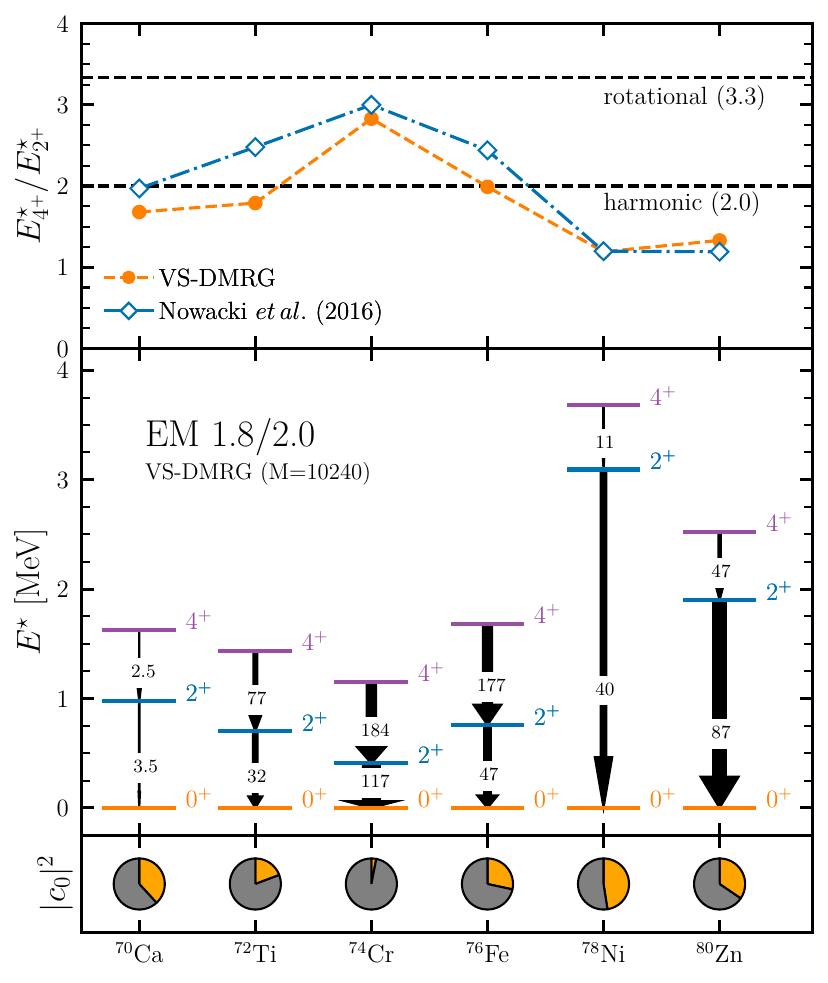}
    \caption{Top panel: Ratio of $4^+$ to $2^+$ excitation energies of $N=50$ isotones for our VS-DMRG results and for the phenomenological shell-model calculations from Nowacki {\it et al.}~\cite{Nowacki2016}, in comparison to a rotational and single-particle-like (``harmonic'') spectrum. Middle panel: Excitation energies and $B(E2)$ transition strengths in units of e$^2$\,fm$^4$ for the lowest $0^+$, $2^+$, and $4^+$ states from our VS-DMRG calculations at bond dimension $M=10240$. Bottom panel: $T_\text{max}=0$ component of the ground-state wave function (orange fraction).}
    \label{fig:N50spec}
\end{figure}

With the performance of the VS-DMRG being validated for spectroscopic observables, we now turn to studying the low-lying spectroscopy of $N=50$ isotones from calcium ($Z=20$) to zinc ($Z=30$).
Due to their neutron-rich character, except for \elem{78}{Ni} and \elem{80}{Zn}, no spectroscopy data is available~\cite{Taniuchi2019}.
Hence we compare our results to state-of-the-art shell-model calculations by Nowacki \etal~\cite{Nowacki2016}.
They observed a rapid transition from single-particle-like excitations in \elem{78}{Ni} to collective rotational excitations in \elem{74}{Cr}.
Rotational structures can be approximately extracted by comparing the level spacings of the low-lying spectrum to that of a rigid rotor, $E^\star (J) \sim  J(J+1)$~\cite{Bohr98nucstruc}.
For the case of the lowest $4^+$ and $2^+$ state the ratio of the energies \Efourptwop{} is given in the top panel of Fig.~\ref{fig:N50spec}. For a perfect rotor the expected value is $\Efourptwop{} = 3.33$, whereas for single-particle-like excitations the ratio will be lower.
Starting from doubly magic \elem{78}{Ni} there is a rapid transition to the rotational spectrum of \elem{74}{Cr} with a ratio of $\Efourptwop{} = 2.83$.
Our VS-DMRG predictions qualitatively match the shell-model results from Ref.~\cite{Nowacki2016} with noticeable deviation in \elem{72}{Ti}. Still there is a clear transition towards the rigid-rotor value in \elem{74}{Cr}, which suggests that our \ai{} calculations qualitatively capture this deformation physics.

Moreover, we show in Fig.~\ref{fig:N50spec} the lowest $0^+$, $2^+$, and $4^+$ states with the allowed $B(E2)$ transitions.
The $2^+$ state in \elem{78}{Ni} at $E^\star \approx 3.0 \, \MeV$ is significantly higher in energy due to the proton shell closure at $Z=28$ and we observe a rapid decrease towards a mid-shell minimum at chromium ($Z=24$) with $E^\star \approx 0.5 \, \MeV$ and another increase towards calcium.
The calculated $B(E2)$ values show a maximum for \elem{74}{Cr} for the $2^+ \rightarrow 0^+$ transition that is characteristic for collective rotational excitations and a signature of nuclear deformation.
In Ref.~\cite{Nowacki2016} this was explicitly confirmed by performing a symmetry-broken and -restored mean-field calculations revealing a pronounced prolate minimum for \elem{74}{Cr} with axial deformation parameter $\beta \approx 0.35$.

We further show the fraction of the $T_\text{max}=0$ component of the ground-state wave function in the bottom panel of Fig.~\ref{fig:N50spec}.
While in the case of \elem{78}{Ni} this component is given by $|c_0|^{2} \approx 0.48$ indicating a dominant occupation of the energetically lowest configurations in the ground state, this changes drastically towards mid-shell \elem{74}{Cr} with very small $|c_0|^{2} \approx 0.03$. Similar observations hold to a lesser extent for the neighboring nuclei \elem{76}{Fe} ($|c_0|^{2} \approx 0.28)$ and \elem{72}{Ti} ($|c_0|^{2} \approx 0.19$).

The general trends of our $B(E2)$ values agree with the shell-model calculations of Ref.~\cite{Nowacki2016} with a significant increase of $B(E2)$ values towards mid-shell chromium -- as expected for a deformed nucleus. The case of \elem{78}{Ni} provides an exception, where our predicted $B(E2)$ value of $40\,e^2\,\text{fm}^4$ is larger than the shell-model result of $34\,e^2\,\text{fm}^4$ for the ground-state transition.
The case of $2^+\rightarrow 0^+$ in \elem{80}{Zn} is the only nucleus with experimental data available and the measured value of $B(E2)=144( 29) \, e^2 \text{fm}^4$~\cite{VanDerWalle2009} is significantly larger than our \ai{} result.
Similarly, the calculated $2^+$ excitation energy in \elem{80}{Zn} of $E_{2^+}^\star = 1.90 \, \MeV$ overpredicts the experimental value of $E_{2^+}^\star = 1.49 \, \MeV$, hinting at an incomplete account of the underlying deformation physics.
We attribute our qualitative agreement of the $B(E2)$ values in \elem{78}{Ni} to the dominance of single-particle excitation whereas the increased differences in mid-shell isotones reveal stronger sensitivity to collective effects.
Still, the size of deformation effects is underpredicted compared to phenomenological shell-model results.
In addition, Nowacki \etal{} further predict an excited-state rotational band with a deformed $0^+_2$ bandhead located around $2.6 \, \MeV$ in agreement with the experiment~\cite{Taniuchi2019}.
In our simulations we were unable to reproduce this deformed state and we find the $0^+_2$ state at a much higher energy of $E^\star \approx 5.0 \, \MeV$ for $M=10240$.
While its convergence as a function of bond dimension is somewhat slower compared to the other states (not shown), the DMRG uncertainty is well under control.
Indeed, previous coupled-cluster calculations using the same NN+3N interaction were also unable to reproduce the deformed excited band~\cite{Hage16Ni78}.

\section{Quantum information analysis of ground states}\label{sec:QIT}

\begin{figure}[t!]
\centering
\includegraphics[width=\columnwidth]{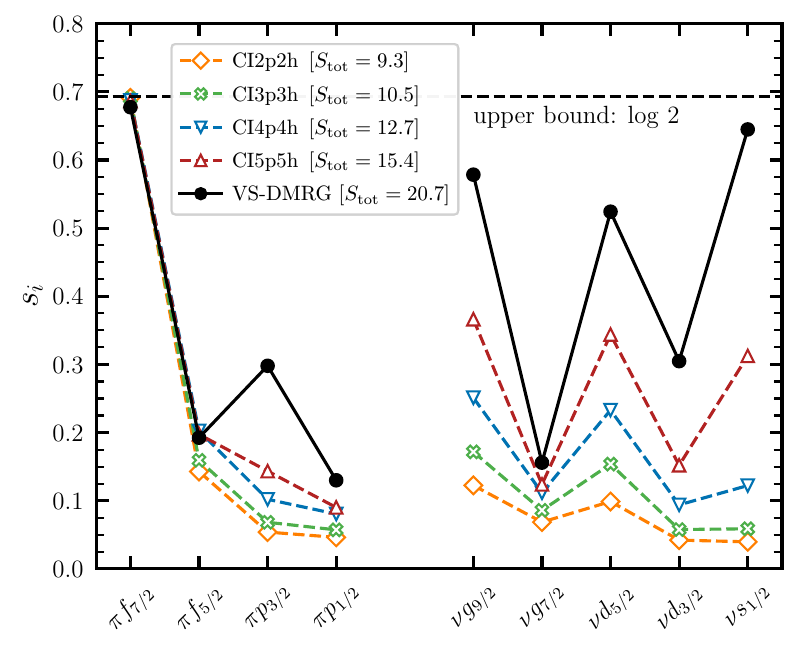}
\caption{Single-orbital entropies for the ground state of \elem{74}{Cr} in the VS-DMRG, in comparison with increasing CI particle-hole truncations. The corresponding total entropies are given in the legend.}
\label{fig:entropy}
\end{figure}

We finally turn to a discussion of orbital entanglement for the strongly correlated \elem{74}{Cr} nucleus using established QI measures.
Many-body correlation effects can be explicitly linked to orbital entanglement between single-particle states for a given nuclear wave function~\cite{Legeza2015}.
We therefore, define the subsystem's reduced density matrix $A \subset \mathcal{H}_1$ via $\rho_{A} = \trace_{B} \, \rho_{AB} $, where $A\cup B = \mathcal{H}_1$. Setting $A = \{p\}$ we introduce the single-orbital entropy as $s_i = - \trace \, \rho \, \log \, \rho$ that is directly linked to the occupation numbers $n_p$ via $s_p = - n_p \, \log  \, n_p - \bar n_p \, \log \, \bar n_p$ where $\bar n_p \equiv 1- n_p$.
Therefore, single-orbital entropies can be straightforwardly evaluated in the DMRG and CI frameworks.
Our previous work showed that the orbital entropies can be linked to the closed-shell character of a nucleus~\cite{Tichai2023dmrg}.

In Fig.~\ref{fig:entropy} we plot the single-orbital entropies for \elem{74}{Cr} for the VS-DMRG wave function and a selection of CI truncations $\Xmax{T}=2,3,4,5$. 
In the legend we further display the total entropy $S_\text{tot} = \sum_p s_p$ for all cases considered.
For most orbitals, the CI-based single-site entropies only slowly converge towards the VS-DMRG values. This effect is particularly strong for the $\nu g_{9/2}$, $\nu d_{5/2}$ and $\nu s_{1/2}$ orbitals and the difference in total entropy can be almost entirely attributed to these three orbitals. It is expected that correlations between these low-lying $\Delta l=2$ orbitals drive quadrupolar deformations.
Compared to \elem{78}{Ni} (not shown) there is enhancement by a factor of three for these orbitals, indicating a strong quadrupole correlation. Along the same lines, the total entropy $S_\text{tot} = 10.4$ in \elem{78}{Ni} is only half the size of the value obtained for \elem{74}{Cr} ($S_\text{tot} = 20.7$), again highlighting the doubly magic character of \elem{78}{Ni}~\cite{Tichai2023dmrg}.
On the contrary, we observe only minor differences in single-orbital entropies between CI and DMRG for selected orbitals ($\pi f_{7/2}$, $\pi f_{5/2}$ and $\nu g_{7/2}$) that indicate that the correlations from these orbitals are relatively well captured in low-rank CI calculations.
Interestingly, for the $\pi f_{7/2}$ orbital already the CI-2p2h agrees with the DMRG data, (almost) fully exhausting the upper limit of $s_p = \log \, 2$ (or equivalently $n_p = 1/2$).
While the general trend of single-site entropies is qualitatively captured at $\Xmax{T} = 5$, the entropy values are generally much lower. In addition, the increase for the $\pi p_{3/2}$ is not reproduced by any of the CI wave functions.
The much increased entropy values reveal that MPS states capture correlations that are absent in particle-hole-truncated CI wave functions and will be of $6$p-$6$h-character and beyond.

\section{Conclusion and outlook}

In this work, we compared the convergence behavior of the VS-DMRG approach for the $N=50$ isotones and compared it to the performance of CI expansions. This showed the improved convergence of the VS-DMRG for ground- and excited-state energies as well as for electromagnetic transitions.
This puts the VS-DMRG approach on par with traditional shell-model machinery, while providing a much more scalable approach to excessively large valence spaces beyond the scope of traditional diagonalization techniques. Our work highlights the role of the tensor topology on the convergence in nuclear physics applications.

Finally our methodology was applied to $N=50$ isotones to obtain an \ai{} description of transitional nuclei at the $N=50$ shell closure. While our predictions qualitatively agree with previous state-of-the-art shell-model calculations, we fail to reproduce the deformed excited-state band in \elem{78}{Ni}, hinting at increased many-body uncertainties or interaction deficiencies.

While the computational demands of the VS-DMRG implementation are tractable in this work, we envision more advanced schemes to cope with computational challenges emerging in larger valence spaces. This can be either achieved by a symmetry-adapted formulation of the DMRG explicitly accounting for non-Abelian SU(2) symmetries (see Ref.~\cite{McCulloch2002}), or by a more careful subspace selection using restricted-active-space formulations, where the coupling to weakly entangled high-energy orbits is approximated by limiting oneself to low-rank CI excitations~\cite{Barcza2022ras,Friesecke2023}.
This work paves the way for accessing configuration spaces well above current shell-model capacities.
These arise for heavy nuclei and nuclei far away from shell closures, and will be particularly important for deriving \ai{} constraints on heavy neutron-rich nuclei.

\section*{Acknowledgements}

The work of A.T., T.M., and A.S.~was supported by the European Research Council (ERC) under the European Union's Horizon 2020 research and innovation programme (Grant Agreement No.~101020842). M.A.W.~was supported by the Bolyai Research Scholarship of the Hungarian Academy of Sciences.
\"O.L.~and G.Z.~were supported by the Quantum Information National Laboratory of Hungary and by the Hungarian National Research Development and Innovation Office (NKFIH) through Grants Nos.~K134983, SNN139581, and TKP2021-NVA-04. 
\"O.L.~received further support from the Hans Fischer Senior Fellowship programme funded by the Technical University of Munich -- Institute for Advanced Study and from the Center for Scalable and Predictive methods for Excitation and Correlated phenomena (SPEC), funded as part of the Computational Chemical Sciences Program FWP 70942 by the U.S.~Department of Energy (DOE), Office of Science, Office of Basic Energy Sciences, Division of Chemical Sciences, Geosciences, and Biosciences at Pacific Northwest National Laboratory.
Computations were in part performed with an allocation of computing resources at the Jülich Supercomputing Center.
We have used the \texttt{NuHamil}~\cite{nuhamil}, \texttt{imsrg++}~\cite{Stro17imsrggit}, and \texttt{KSHELL}~\cite{kshell} codes
for the 1.8/2.0 (EM) NN+3N matrix elements generation, VS-IMSRG decoupling, and CI diagonalization, respectively.

\bibliographystyle{apsrev4-1}
\bibliography{strongint}

\end{document}